\documentclass[prl,amsmath,aps,twocolumn,superscriptaddress]{revtex4}
\usepackage{amsmath}
\usepackage{amssymb}
\usepackage{epsfig}
\usepackage{amscd}
\usepackage{graphicx,psfrag,xspace}
\usepackage{float}
\usepackage[normalem]{ulem}

\begin{document}

\title{First passage times and distances along critical curves}

\author{A. Zoia}
\email{andrea.zoia@polimi.it}
\affiliation{Department of Physics, Massachusetts Institute of Technology, Cambridge, Massachusetts 02139, USA}
\affiliation{Department of Nuclear Engineering, Polytechnic of Milan, Milan 20133, Italy}
\author{Y. Kantor}
\affiliation{School for Physics and Astronomy, Raymond and Beverly Sackler Faculty of Exact Sciences, Tel Aviv University, Tel Aviv 69978, Israel}
\author{M. Kardar}
\affiliation{Department of Physics, Massachusetts Institute of Technology, Cambridge, Massachusetts 02139, USA}

\begin{abstract}
We propose a model for anomalous transport in inhomogeneous environments, such as fractured rocks, in which particles move only along pre-existing self-similar curves (cracks). The stochastic Loewner equation is used to efficiently generate such curves with tunable fractal dimension $d_f$. We numerically compute the probability of first passage (in length or time) from one point on the edge of the semi-infinite plane to any point on the semi-circle of radius $R$. The scaled probability distributions have a variance which increases with $d_f$, a non-monotonic skewness, and tails that decay faster than a simple exponential. The latter is in sharp contrast to predictions based on fractional dynamics and provides an experimental signature for our model.
\end{abstract}
\maketitle

\section{Introduction}

Scale invariant curves, such as coastlines and fracture fronts, abound in nature \cite{mandel_book}. The fractal dimension $d_f$ relates the scaling of the length $\ell$ along the curve between two points to their actual separation $R$ in Euclidean space by $\ell\propto R^{d_f}$. However, for stochastic shapes, the actual length will fluctuate, and knowledge of its full distribution greatly augments the information from scaling. We numerically investigate some related distributions for scale invariant (critical) curves in two dimensions.

While we measure the distribution of lengths along fractal curves, an underlying motivation is to understand the distribution of times for traversing such curves. In particular, we are interested in exploring flow through porous media, such as diffusion of chemical tracers injected into a water-saturated heterogeneous permeable sediment or fractured hard rock formation. Experiments suggest that flow proceeds along a network of pre-existing fractal paths \cite{berkowitz1}, as further corroborated by simulations modelling streams as percolation clusters \cite{kimmich}. In such cases, finding the distribution of first passage times between points of known Euclidean distance requires knowledge of the dynamics of the process, as well as the distribution of traversed lengths. If we assume that the motion along the path is independent of its shape (as in a simple damage spreading process, or in curvature independent diffusion), we obtain the distribution of first passage times by simple manipulation of the distribution of lengths. At the minimum, results based on such an assumption help to rule out simple hypotheses about the dynamics, and/or structures of the fractal shapes, in physical models.

Important examples of scale invariant curves in two dimensions are self-avoiding walks (SAWs) and percolation fronts (PFs); additional instances may be obtained from interacting systems at a critical point \cite{madras, cardy_perc}. These examples suggest that generating and studying stochastic scale-invariant curves is a computationally hard process. Indeed, the past decades have witnessed much effort and progress in efficient numerical algorithms for generating SAWs and critical systems \cite{landau}. An important recent development concerns the Stochastic Loewner Evolution (SLE), which provides a way of generating such curves through mappings (in the complex plane) of a simple Markovian random walk \cite{werner0, cardy, gruzberg}. While the main interest in SLE has been to extract analytic information about critical systems, here we use it as an efficient means to generate curves with tunable fractal dimension. 

\section{SLE, critical curves and their lengths}

Specifically, we study non-self-intersecting curves in two-dimensions that begin at the origin and remain in the upper half-plane. Such curves can be parameterized  by a complex function $\gamma(t)$ of a real argument $t$. Loewner has shown that for any such curve described by $\gamma(t')$, with $0 \le t' \le t$, there exists a conformal mapping $g_t(z)$ that maps the upper half-plane surrounding the curve into the complete half-plane, while the curve itself is mapped onto the real axis \cite{loewner}. In particular, for the tip of the curve $g_t(\gamma(t))\equiv\zeta(t)$, where $\zeta(t)$ is a real function. According to Loewner, the mapping $g_t$ satisfies the equation
\begin{equation}
\frac{dg_t(z)}{dt}=\frac{2}{g_t(z)-\zeta(t)}.
\label{sle}
\end{equation}
This equation has a natural parameterization in terms of the so-called Loewner time $t$, which has dimensions of an area and whose definition is based on a standard form of $g_t$ \cite{cardy}. The knowledge of the real function $\zeta(t)$ enables a complete reconstruction of the curve $\gamma(t)$: this is accomplished by integration of Eq.~(\ref{sle}) backwards in time, starting from the ``final condition" $g_t(\gamma(t))=\zeta(t)$ and using $g_{t=0}(z)=z$.

\begin{figure}[t]
\centerline{ \epsfclipon \epsfxsize=9.0cm 
\epsfbox{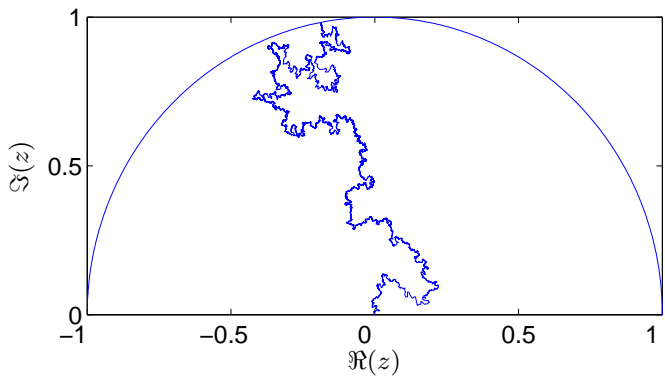} }
   \caption{A discretized implementation of SLE$_{8/3}$, which in the scaling limit is conjectured to be a SAW in the upper half-plane. The curve is generated by integrating Eq.~(\ref{sle}) backward in time with step size $dt=10^{-5}$. The trace $\gamma(t)$ is stopped upon its {\em first encounter} with the semicircle of radius $R=1$.}
   \label{sle_trace}
\end{figure}

It has been shown \cite{schramm0, rohde, werner0, cardy, gruzberg} that different sets of scale invariant curves with fractal dimensions $1< d_f < 2$ map onto standard one-dimensional Brownian motion functions $\zeta(t)=\sqrt{\kappa} B(t)$, with zero mean and with a variance $\kappa$ related to the fractal dimension by $d_f=1+\kappa/8$ \cite{beffara1}. This process is denoted as SLE$_\kappa$ and can be used to conveniently generate long scale--invariant curves. For particular values of $\kappa$ the traces $\gamma(t)$ coincide with the continuum limit of well known curves, such as SAW or PF, defined on critical lattice models \cite{cardy}. An example curve generated by implementing SLE$_{8/3}$ (corresponding to a SAW) with discrete time steps is shown in Fig.~\ref{sle_trace}.

For SLE$_\kappa$ traces such as in Fig.~\ref{sle_trace}, an important issue is how to measure length between two points on the curve. In fact, continuous-time Brownian motion generates SLE$_\kappa$ paths which contain infinitely minute details, thus having an infinite length. This problem has been addressed by Kennedy \cite{kennedy_param, kennedy2006}, and like him we introduce a specific procedure for calculating the lengths of curves generated by time-step discretization. Loewner time is discretized into $m=t/dt$ equal time steps. While the discrete time points $i dt$ are equally spaced, Loewner equation produces a discrete set of positions $z(i dt)$, where the two-dimensional distances between sequential points $\Delta_i=|z(i dt)-z((i-1)dt)|$ are extremely non-uniform. For the solution of the physical problem we need to know the length of the walk between such points. In the lattice version of the walk, we expect the number of steps between two points to be proportional to $\Delta^{d_f}$. One can use this relation to define the length $\ell$ of the walk as
\begin{equation}
\ell \equiv c \sum_{i=1}^m |z(i dt)-z((i-1)dt)|^{d_f},
\label{time}
\end{equation}
with a proportionality factor $c$ with dimensions of $[t]^{(1-d_f)/2}$.

Consistency and suitability of this definition for the standard relations between the number of steps and the end-to-end distance of the curve and its proper behavior in the $m \to \infty$ limit have been demonstrated {\em numerically} by Kennedy \cite{kennedy_param, kennedy2006}. As $dt \to 0$, more points are sampled, their separations $\Delta_i$ become smaller, but the sum in Eq.~(\ref{time}) converges to a {\em finite} limit. Since we use the definition in the context of first passage problems, applied to a stochastic variable, $\ell$, which is different from those considered in \cite{kennedy_param, kennedy2006}, it is essential to verify the correctness of the procedure. We do so by comparing distributions obtained from lattice implementation of SAWs and PFs with those obtained from SLE$_\kappa$ at corresponding values of $\kappa$.

\begin{figure}
\centerline{ \epsfclipon \epsfxsize=9.0cm 
\epsfbox{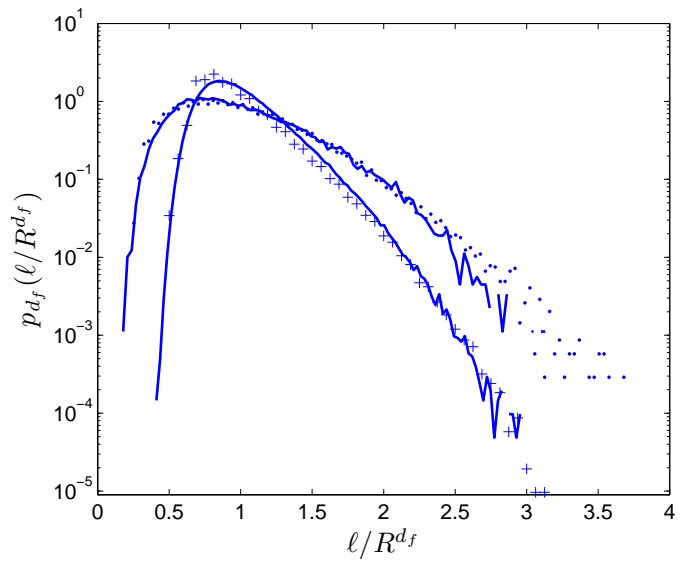} }
  \caption{Comparison between the rescaled distributions of lengths $p_{d_f}(\ell/R^{d_f})$ for SLE$_\kappa$ and for the corresponding lattice models. The solid line with a narrow peak corresponds to the probability distribution of SLE with $\kappa=8/3$ and is compared with that of SAWs, plotted in crosses. The solid line with a wide peak corresponds to the probability distribution of SLE with $\kappa=6$ and is compared with that of PFs, plotted in dots. All distributions have been further rescaled so that their average is equal to one.}
 \label{fig2}
\end{figure}

\section{Distributions of first passage lengths}

Relying upon Eq.~(\ref{time}), we have computed the distribution of lengths for SLE$_\kappa$ with $\kappa=8/3$ and $\kappa=6$ for curves starting from the origin of the upper half-plane and arrested upon touching for the {\em first time} a semicircle of radius $R$. For these values of $\kappa$, SLE$_\kappa$ is conjectured and shown to represent the scaling limit of SAWs on a square lattice and PFs on a triangular lattice, respectively \cite{schramm, kennedy_saw_sle, smirnov}. We have computed via Monte Carlo simulation the length distributions for these lattice models (which have a natural definition of length), adopting the same boundary conditions. In Fig.~\ref{fig2} we compare the rescaled $p_{d_f}(\ell/R^{d_f})$ for the two sets of distributions: the numerical value of $c$ in Eq.~\ref{time} for SLE$_\kappa$ is fixed by imposing that the average of the distribution is one.

The agreement between SLE$_\kappa$ and the corresponding lattice models is quite good, thus supporting our choice of Eq.~(\ref{time}) as a proper means of measuring lengths. The quantity which is calculated as $\ell$ is therefore the true length (in the usual sense) of a polymer chain for $\kappa=8/3$, of a percolation front for $\kappa=6$, or in general of a scale invariant self-avoiding curve. The somewhat unintuitive form of Eq.~(\ref{time}) is a consequence of Loewner evolution, which, when used as a finite difference equation with equally spaced time instances $t_i$, produces a non-uniform distribution of Euclidean distances $z(t_i+dt)-z(t_i)$ \cite{kennedy2006}.

However, we should stress that there are also notable differences between the two sets of curves: one possible source of discrepancies comes from finite-size (lattice) effects, since SLE$_\kappa$ corresponds to a continuum limit in which the number of steps on a lattice goes to infinity, while the lattice spacing goes to zero. A second source is intrinsic to SLE$_\kappa$ for $\kappa> 4$. At these values the traces can touch themselves, the numerical integration of Eq.~(\ref{sle}) becomes problematic \cite{kennedy2006}, and convergence to the asymptotic limit is slow.

\begin{figure}[t]
\centerline{ \epsfclipon \epsfxsize=9.0cm 
\epsfbox{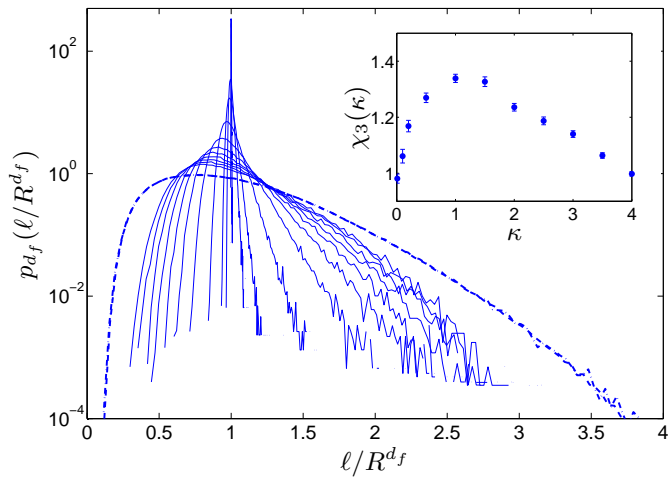} }
   \caption{Rescaled distributions of lengths for SLE$_\kappa$ with $\kappa=0.01, 0.1, 0.2, 0.5, 1, 1.5, 2, 2.5, 3, 3.5$ and $4$. The distributions have been further rescaled so that their average is one. As $\kappa$ ($d_f$) increases, the peak of the distribution moves to the left while its width becomes larger. Dashed lines represent the rescaled distributions for PFs (corresponding to $\kappa=6$) for $R=100$ and $R=500$. It should be noted that the two distributions are almost indistinguishable, and clearly show a faster than exponential decay. The inset shows the skewness, $\chi_3$, of the length distributions as a function of $\kappa$.}
   \label{sle_curves_inset}
\end{figure}

Having gained confidence in the validity of our numerical measure of length along critical curves, we use SLE$_\kappa$ to generate curves for a range of $d_f$ for which there is no simple lattice analog.  It is worth emphasizing that the appearance of SLE$_\kappa$ enables dealing with problems which were previously untreatable: although SAWs and PFs can also be calculated directly, most other scale-invariant curves with a generic $d_f$ cannot be produced by any other method. The resulting scaled probability distributions of the lengths of these curves, $p_{d_f}(\ell/R^{d_f})$, are plotted in Fig.~\ref{sle_curves_inset}. Each distribution is obtained from $10^{5}$ SLE$_\kappa$ trajectories starting from the origin and arrested upon first touching a semicircle with radius $R$. The integration step for all curves is $dt=5\times 10^{-4}$. The distributions depend only on the scaled parameter $\ell/R^{d_f}$, and we have further rescaled both axes such that the mean value is at unity. The distributions are asymmetric and become broader as $d_f$ is increased. The variance initially grows quadratically in $\kappa$, then levels off to a slower growth. Interestingly, the skewness (a measure of asymmetry) is not monotonic in $d_f$ and has a maximum at $\kappa \approx 1$, as depicted in the inset of Fig.~\ref{sle_curves_inset}. The entire shape of the distribution (and by extension the variance and the skewness) are universal functions of $d_f$.

For comparison with other systems, an important question concerns how the distribution behaves at large arguments. Detailed numerical simulations of PFs allow us to conclude that the tails of the distributions decay faster than exponentially. We generated PFs for the relatively large sizes of $R=100$ and $R=500$ ($10^{7}$ and $10^{6}$ trajectories, respectively). As indicated in Fig.~\ref{sle_curves_inset}, a non-linear behavior at large arguments is present in this semi-log plot, irrespective of $R$ (implying that finite-size effects can be neglected). Intuitively, we expect this feature to arise from the self-avoiding nature of SLE$_\kappa$: for a Markovian path allowed to intersect itself, the probability of adding a new segment to an existing trace in the domain is independent of its length. Such paths are characterized by a distribution that decays as a simple exponential. For SLE$_\kappa$, by contrast, the available space within the domain decreases with increasing $t$ ($\ell$) and the escape probability increases.

\section{Application to transport in porous media}

Our results lead to a practical application in the context of {\em anomalous transport}: as a specific example, we focus on the paradigmatic case of flow in rocks, in which motion is constrained to proceed along preexisting (fractal) curves. Non-diffusive dynamics is quite common in inhomogeneous environments such as these, and frequently modeled by a fractional Fokker-Planck equation (FFPE) \cite{klafter1, klafter2}. In the FFPE approach, the medium where particles move is supposed to be globally homogeneous, and the effects of heterogeneities are taken into account by assuming that particles are locally trapped and thereby experience extremely long (power-law distributed) waiting times. This is then mathematically modelled through fractional derivatives in time. However, the behavior resulting from specific physical realizations (such as motion along scale invariant curves) can be quite different from that arising from ideal fractional dynamics. To elucidate these differences, we focus on the first passage time distribution (FPTD), namely the probability that a walker starting at the origin in Fig.~\ref{sle_trace} reaches the circle at radius $R$ at time $\tau$ {\em for the first time}. 

Within the FFPE approach, the trapping events lead to the anomalous transport scaling $\tau\propto R^z$, with $z>2$. Analysis of the equation then indicates that for a domain of finite size the FPTD has an asymptotic power-law decay as $\tau^{-1-2/z}$ for times much longer than the typical time scale of the process (the specific nature of the boundaries does not affect this scaling) \cite{klafter2}\footnote{See Eq.~(66) of Ref.~\cite{klafter2}, and note that $\alpha=2/z$ in our notation.}. FFPE also allows incorporating an additional external drift velocity $v$, and in this case the FTPD (after an initial transient whose duration is proportional to the ratio $R/v$) decays as $\tau^{-1-1/z'}$, with $z'>1$ \cite{barkai}\footnote{See Sec. V of Ref.~\cite{barkai} (Scher-Montroll transport): for the biased case note that $R^2 \sim \tau^{2 \alpha}$, and our scaling for the FPTD asymptotics can be easily recovered.}. In both cases, the long retention times are such that {\em the mean time to reach the outer boundary is infinite} \cite{klafter2}.

In our model of flow along scale-invariant streams, the time it takes to first reach a Euclidean distance $R$ depends on two elements. One is the dynamics by which motion proceeds along these tracks, and the second is the distribution of lengths $\ell$ of appropriate tracks for a given Euclidean distance $R$. The latter is the quantity that we calculated in Fig.~\ref{sle_curves_inset}; the former can depend on many features of the curve (such as curvature, vicinity of other portions) as well as on its length. To make the problem tractable, we make the simple assumption that the probability that the walker reaches its destination (for the first time) at a time $\tau$ only depends on the length $\ell$, and is given by some function $p(\tau | \ell)$. The overall FPTD is then obtained by the integral $p_{d_f}(\tau|R)=\int d \ell p(\tau|\ell) p_{d_f}(\ell|R)$.

\begin{figure}[t]
\centerline{ \epsfclipon \epsfxsize=9.0cm 
\epsfbox{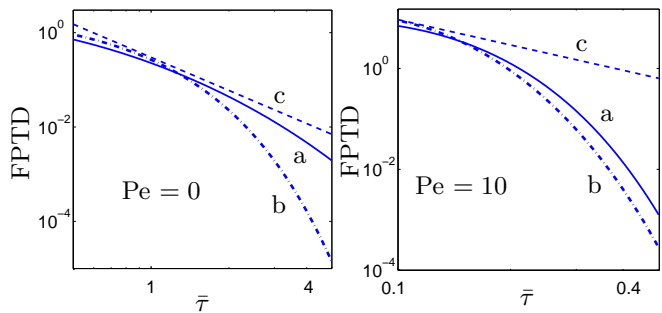} }
   \caption{FPTD along critical curves with $d_f=1.5$ (a); contrasted with the results for normal diffusion (b); and FFPE (c). Left: diffusion-dominated transport (Pe=0). Right: drift-diffusion (Pe=10). Time is in dimensionless units $\bar{\tau}=\tau D/R^2$. Note that for (c) the mean first passage time is infinite.}
   \label{pdf_times}
\end{figure}

In a damage spreading process (such as burning fire) in which the walker proceeds at a constant velocity, $p(\tau|\ell)=\delta(v\tau-\ell)$. More generally, for a drift-diffusion process we must solve the eigenvalue problem for first passage over a segment of length $\ell$, which depends on the P\'eclet number Pe$=v \ell /D$ and on the initial and boundary conditions on the interval \cite{redner}. To be consistent with the macroscopic boundary conditions on the semicircle, we impose that the particles are injected at the origin and can escape from the track only upon touching the other extreme of the curve at the outer boundary of the traversed medium, where they are absorbed. Mathematically, this is implemented by assigning a reflecting boundary condition at the origin (any particles moving back to the injection point are sent back without loss), and an absorbing boundary on the end point at $\ell$. Under these assumptions, the first passage time distribution reads as follows:
\begin{equation}
p(\tau|\ell)=2e^{\zeta} \sum_j\frac{e^{-(\lambda_j^2+\zeta^2)\tau}}{1+\frac{\zeta}{\lambda_j^2+\zeta^2}}\lambda_j \sin(\lambda_j),
\label{fptd_pdf}
\end{equation}
where $\lambda_j$ are the roots of $\tan(\lambda)=-\lambda/\zeta$, and $\zeta=\rm{Pe}/2$ \cite{hinkel}.

In Fig.~\ref{pdf_times} we plot the resulting FPTD, $p_{d_f}(\tau|R)$, for particles diffusing along SLE$_\kappa$ (obtained as a numerical convolution). The left figure corresponds to pure diffusion (Pe $=0$) along the curve, which appears sub-diffusive in Euclidean space, since $\tau\propto\ell^2\propto R^{2d_f}$ and thus $z=2d_f>2$. The results are contrasted in the same figure with normal diffusion (exponentially decaying FPTD), and with that of FFPE (power-law decay). FPTDs for finite values of Pe are plotted on the right side, which appear as super-diffusion in real space, as $\tau\propto\ell\propto R^{d_f}$, whence $z'=d_f<2$. The corresponding curves for normal diffusion and FFPE are also shown. As discussed before, when motion along the curve proceeds at uniform velocity (no diffusion and Pe $\rightarrow \infty$) the tails of the distribution decay faster than a simple exponential. Thus, in all cases the presence of a fairly wide distribution of path lengths broadens the FPTD, but the tails of the distribution fall off more rapidly than predicted by fractional dynamics: this allows for a very strong discrimination between other self-intersecting and non-self-intersecting trajectories at microscopic level. (In particular the MFPT is {\em finite}.) This difference is not detectable by just looking at the macroscopic anomalous transport scaling exponent. Moreover, since FPTD is usually an easily experimentally accessible quantity, this observation may provide a way to support or reject the hypotheses about the microscopic dynamics which lie at the basis of these models.

\section{Conclusions}

In this Letter we have proposed a model for anomalous transport in inhomogeneous environments, which is suitable to capture, e.g., the dynamics of flow streams through fractured rocks. In particular, the two-dimensional probability density
of the particles which can be obtained from our model qualitatively agrees with the typical distribution of a chemical plume flowing through a two-dimensional fracture network (as shown in Fig. 4(b) of Ref. \cite{berkowitz_frac}).

Unlike fractional dynamics, the FPTD does not fall off as a power law and has a finite mean. A necessary element of the model is the distribution of lengths of fractal curves, which we obtain by numerical studies of lattice models and SLE: to the best of our knowledge, the first passage distributions obtained here have not been derived analytically. We hope that analytical approaches to SLE can provide exact insights on these distributions, and hence enable quantitative comparisons with experiments.

\acknowledgments
This work was supported by the NSF grant DMR-04-2667 (M.K.). A.Z. thanks the Fondazione Fratelli Rocca for support through a Progetto Rocca fellowship.

\end{document}